\documentclass[journal]{IEEEtran}
\ifCLASSINFOpdf
\usepackage[pdftex]{graphicx}
\else
\fi
\usepackage{amsfonts}
\usepackage{amsmath}

 \usepackage{multirow}
 \usepackage{booktabs}
 \usepackage{tabularx}
 \usepackage{enumerate}
 \usepackage{color}
 \usepackage{cite}
\newcommand{\e}[1]{\ensuremath{\times 10^{#1}}}

\begin{document}
%
\title{Blind SAR Image Despeckling Using Self-Supervised Dense Dilated Convolutional Neural Network}
%
%
%
\author{Ye~Yuan, Jian~Guan*, Jianguo~Sun
\thanks{Y. Yuan, J. Guan and J. Sun are with the College of Computer Science and Technology, 
Harbin Engineering University, Harbin 150001, China (Corresponding author: J. Guan, e-mail: j.guan@hrbeu.edu.cn).}}
\maketitle

\begin{abstract}
Despeckling is a key and indispensable step in SAR image preprocessing, existing deep learning-based methods achieve SAR despeckling by learning some mappings between speckled (different looks) and clean images. However, there exist no clean SAR image in the real world. To this end, in this paper, we propose a self-supervised dense dilated convolutional neural network (BDSS) for blind SAR image despeckling. Proposed BDSS can still learn to suppress speckle noise without clean ground truth by optimized for L2 loss. Besides, three enhanced dense blocks with dilated convolution are employed to improve network performance. The synthetic and real-data experiments demonstrate that proposed BDSS can achieve despeckling effectively while maintaining well features such as edges, point targets, and radiometric. At last, we demonstrate that our proposed BDSS can achieve blind despeckling excellently, i.e., do not need to care about the number of looks.
\end{abstract}


%
\IEEEpeerreviewmaketitle

\section{Introduction}
%
%
%
%
\IEEEPARstart{S}{ynthetic} aperture radar (SAR) is an active image remote sensing system that transmits the microwave signals in a side looking direction towards the earths' surface. The benefit of SAR imaging systems over optical includes high-resolution images independent of daylight and weather-related occlusions i.e., clouds, dust and snow \cite{fetterer_sea_1994,bovolo_hierarchical_2013}. This all-day and all-weather acquisition capability sometimes makes the SAR imaging system reliable than optical imaging system for analyzing earth resources, such as ground and sea monitoring, disaster assessment, and so on \cite{touzi_review_2002}. 

However, a major problem of SAR image is the presence of speckle \cite{gao_statistical_2010}. Speckle is a granular noise caused by constructive or destructive interference of backscattered microwave signals which results in the obtained active image quality. It strongly impairs the performance of the aforementioned tasks. Thus, speckle removal is a key and indispensable step in SAR image preprocessing.

To removal the speckle from SAR image, this called for an intense research activity on SAR despeckling in the past decade \cite{argenti_tutorial_2013}. Overall, not absolutely, existing despeckling methods can be grouped into three categories, local window filters, non-local means (NLM)-based methods, and convolutional neural network (CNN)-based methods. 

The most commonly applied filtering technique restores the center pixel in a moving window with an average of all the pixels in the windows. In early years, scholars have proposed a lot of despeckling methods based local window filtering, such as, Lee filter \cite{lee_digital_1980}, Refined Lee filter \cite{jong-sen_l_scattering-model-based_2006}, Frost filter \cite{frost_model_1982}, Kuan filter \cite{kuan_adaptive_1987}, and Gamma-MAP filter \cite{lopes_structure_1993}. These filters can smooth the speckle noise, however, different sizes of window and the number of looks can result in varying filtering quality, as the size of filter window and the number of looks increase, the more speckle noise is decreased and edge information is lost.

In order to overcome the deficiency of local window filters, the NLM-based methods have been applied to process SAR image in recent years \cite{chen_nonlocal_2011, hua_zhong_robust_2014, liu_nonlocal_2014,deledalle_nl_sar_2015}. In the NLM-based methods, filtering is carried out, through the weighted mean of all the pixels in a certain neighborhood. However, differing from the local window filters, the weight does not depend on the geometric distance between the target pixel and each given pixel, but on their similarity. The similarity is measured by the Euclidean distance between the patches surrounding the selected and the target pixels. This principle has inspired several methods, such as BM3D \cite{dabov_image_2007}, where the nonlocal approach is combined with wavelet shrinkage and Wiener filtering in a two-step process. And then, aimed to taking into account the SAR despeckling peculiarities, SAR-BM3D \cite{parrilli_nonlocal_2012}, a SAR-oriented version of BM3D is proposed.

Local window filters and NLM-based methods both have the intensive computational complexity and burden. This may hinder their usage in practical applications. Other than this, for a given SAR image, these methods require the number of looks known. In other words, they can not achieve ``Blind Despeckling". However, in many cases, the number of looks in a SAR image is unavailable due to the uncertainties of sensors.

Recently, deep learning has been a hot topic in image processing over the last decade \cite{lecun_deep_2015}. Benefiting from the new thoughts and theories in this area, CNN-based despeckling methods have gradually emerged in recent years ~\cite{wang_sar_2017, gu_residual_2017, chierchia_sar_2017, kim_despeckling_2018, zhang_learning_2018, tang_sar_2018, gui_sar_2018}. Their basic goal is to model the nonlinear relationship between speckle and clean images, i.e., supervised learning. This happens by training CNNs with numerous pairs of noisy inputs and clean targets. Differing from nature image denoising task, the significant problem of SAR despeckling is that there exist no clean targets in the real world. SAR-CNN \cite{chierchia_sar_2017} solve this problem by using multi-temporal SAR data of the same scene, where original data is used as noisy inputs and corresponding multi-look data as clean targets. Since it is not completely reliable to treat the multi-look images as clean targets, the despeckling effect of SAR-CNN is very limited. Differing from SAR-CNN, SAR-IDN \cite{wang_sar_2017} and SAR-DRN \cite{zhang_learning_2018} use optical photographs as simulated clean SAR images and artificially add speckle as noisy images. In order to achieve blind despeckling, the approach of CNN-based methods is adding speckle noise of different levels to training data. Recent work, Noise2Noise \cite{Lehtinen2018Noise2NoiseLI}, demonstrate that clean targets are unnecessary for denoising task. Under some certain distributional assumptions, such as additive Gaussian noise, Poisson noise, and so on, CNNs can learn to predict the clean signal by training on pairs of independent noisy measurements of the same target. Although Noise2Noise only has noisy images, it can achieve the same well or even better results as if using clean targets. This kind approach of self-supervised learning bring some new ideas for SAR image despeckling. 

In this paper, we propose a framework for blind despeckling of SAR image. The main contributions can be summarized as follows:
\begin{itemize}
	\item[1)] We propose a new state of the art for SAR image blind despeckling, which employs a self-supervised learning approach, i.e., no clean targets are used in the training process. Since the statistical distribution of speckle noise satisfies the unit mean, BDSS can still learn to suppress speckle noise by optimized for $L2$ loss when speckled images are used as ground truth. The noise level of speckle, i.e., the look of SAR images, is not cared about. In other words, BDSS can achieve blind despeckling effectively.	
	\item[2)] The structure of BDSS mainly consists of three enhanced dense blocks, where dilated convolution is used for enlarging the field of view rather than common convolution. In addition, to improve despeckling ability and reduce computational complexity and memory usage, we remove batch normalization layers and replace ReLU with ParametricReLU from each dense block. 
	\item[3)] We confirm PSNR and SSIM on synthetic specke images are the new state of the art. In addition, we set a despeckling experiment on real SAR images from four different sensors. Compared with other methods in terms of visual results and quantitative evaluation indexes, BDSS can remain better image feature, such as edges, point targets, and radiometric while removing speckle noise effectively.
\end{itemize}
 
 The rest of this paper is organized as follows: In Section \uppercase\expandafter{\romannumeral2}, the self-supervised blind despeckling theory and the structure of proposed BDSS model are described.  In Section \uppercase\expandafter{\romannumeral3}, the synthetic and real-data experimental results and analysis are presented. Finally, our conclusion is given in Section \uppercase\expandafter{\romannumeral4}.
 
\section{Methodology}
\subsection{SAR Speckle Noise Degradation Model}\label{sub:SAR Speckle}
Let $y \in \mathbb{R}^{W \times H}$ be the observed signal, $x \in \mathbb{R}^{W \times H}$ be the original signal (e.g., original data), and $n \in \mathbb{R}^{W \times H}$ be the uncorrelated multiplicative speckle. Then assuming that the a SAR image is an average of $L$ looks, the observed signal $y$ is related to $x$ by the following multiplicative model \cite{lee1989noise}:
\begin{equation}
y=nx.
\end{equation}

It is well-known that, for a SAR image, $n$ follows a Gamma distribution and has the following probability density function:
\begin{equation}
p(n)=\frac{1}{\Gamma(L)} L^{L} n^{L-1} e^{-L n},
\end{equation}
where $\Gamma$ is the Gamma function with unit mean and variance $\frac{1}{L}$, and $n>=0$, $L>=1$. Hence, the process of SAR image despeckling is to estimate the original signal (clean images) $x$ from the observed signal (speckled images) $y$.
\subsection{Self-Supervised Blind Despeckling Theory}

\begin{figure*}[htb] 
	\centering 
	\includegraphics{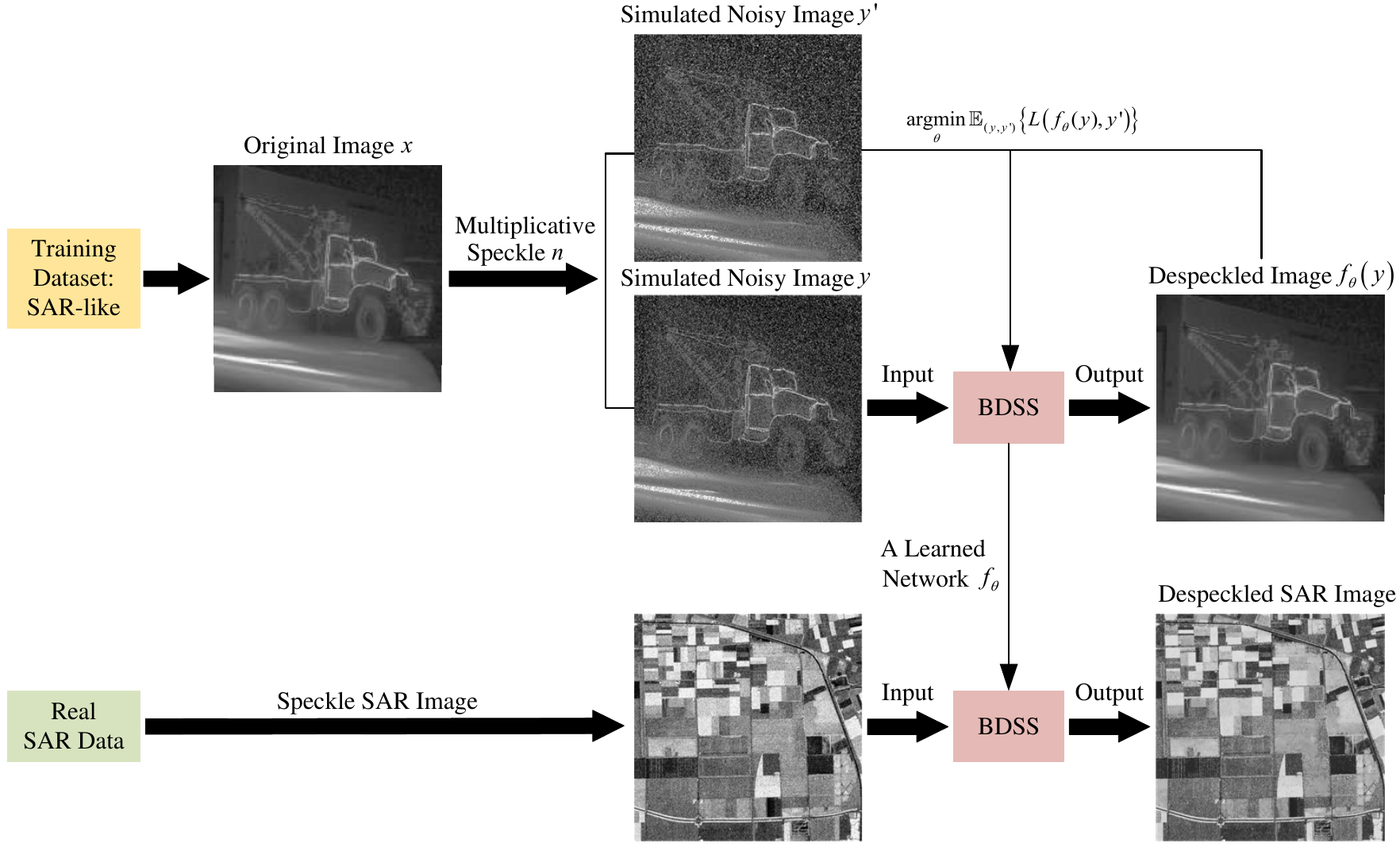} 
	\caption{Flowchart of the proposed BDSS for blind despeckling}
	\label{fig.Flowchart}
\end{figure*}

Assume that we have a set of speckle measurements $\left( {{{y}_1},{{y}_2}, \ldots } \right)$ of the SAR data, in existed CNN-based supervised methods, such as SAR-CNN, ID-CNN and SAR-DRN, a set of training pairs $\left( {{y_i},{x_i}} \right)$ are used to minimize the network loss function in which ${{y_i}}$ is the noisy input image and ${{x_i}}$ is the corresponding clean target image. The network function ${{f_\theta }(x)}$ is parameterized by $\theta $:
\begin{equation} 
\underset{\theta}{\operatorname{arg min}} \mathbb{E}_{(y, x)}\left\{L\left(f_{\theta}(y), x\right)\right\},
\label{equ:FS}
\end{equation}
where $L$ denotes a loss function, normally the ${{L_2}}$ loss $L({f_\theta }(y),x) = {({f_\theta }(y) - x)^2}$.
The ${\theta}$ is found when  ${f_\theta }(y)$ has the smallest average deviation from the $x$. For the ${{L_2}}$ loss, the minimum of Eq.~(\ref{equ:FS}) is found at:
\begin{equation}
{f_\theta }\left( \hat y \right) = x.
\label{equ:FSMIN}
\end{equation}

And then, in our introduced self-supervised SAR despeckling approach, we replace the clean targets ${x_i}$ with corresponding speckle measurement ${y'_i}$, as illustrated in Fig.~\ref{fig.Flowchart}. The network function  Eq.~(\ref{equ:FS})  becomes:
\begin{equation}
\mathop {{\mathop{\rm argmin}\nolimits} }\limits_\theta  {{\mathop{\mathbb{E}}\nolimits} _{(y,y')}}\left\{ {L\left( {{f_\theta }(y),{y'}} \right)} \right\},
\label{equ:SS}
\end{equation}
where $y'$ denotes another speckle SAR image, and both the inputs $y$ and the targets $y'$ are drawn from a corrupted distribution (not the same) conditioned on the underlying. 
For the ${{L_2}}$ loss, the minimum of Eq.~(\ref{equ:SS}) is found at the arithmetic mean of the observations:
\begin{equation}
{f_\theta }\left( {y} \right) = {\mathbb{E}}\{y'\}.
\label{equ:SSMIN}
\end{equation}

As described with Section~\ref{sub:SAR Speckle}, multiplicative speckle noise in a SAR image satisfies unit mean distribution, the arithmetic mean of the observations is the same as clean target:
\begin{equation}
{\mathbb{E}}\{ y'\}  = x,
\end{equation}
this means that Eq.~(\ref{equ:FSMIN}) and Eq.~(\ref{equ:SSMIN})  are equivalent. In other words, the optimal network parameters $\theta $ remain unchanged.

In addition, noting that when the clean targets are replaced with speckle targets whose expected value is the same as the former, what the network learns will not be changed. The noise level of speckle, i.e., the number of looks, is not cared about. This means that the proposed BDSS can achieve reliable blind despeckling.

\subsection{Network Architecture}

\begin{figure*}[htb] 
	\centering 
	\includegraphics{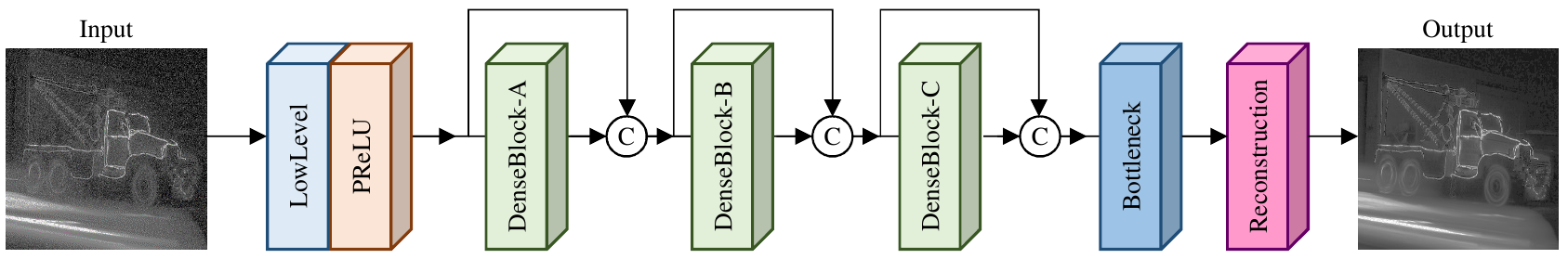} 
	\caption{Structure of BDSS}
	
	\label{fig.BDSS framework}
\end{figure*}

\begin{figure*}[tb] 
	\centering
	\includegraphics{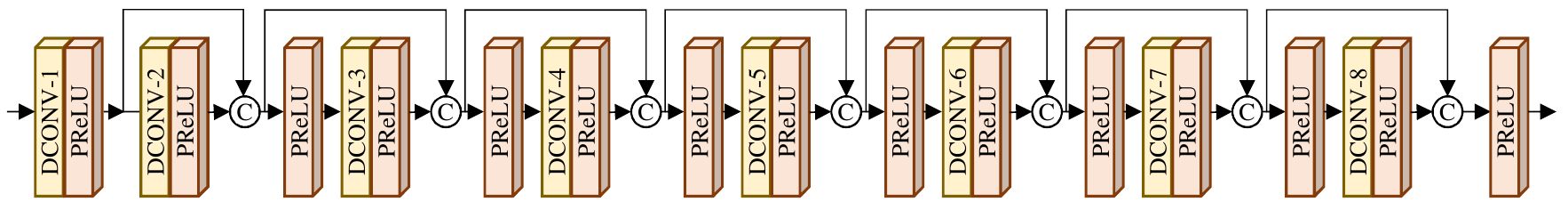} 
	\caption{Structure of enhanced dense block}
	\label{fig.Structure_2}
\end{figure*}

\begin{table}[]	
	\renewcommand{\arraystretch}{1.3}
	\caption{Detailed Configuration of BDSS}
	\label{Configuration}
	\centering
	\begin{tabular}{m{1.7cm}<{\centering}    m{0.4cm}<{\centering}          m{5.4cm}<{\centering}}
		\toprule[2pt]
		Name                           & ${N_{out}}$ & Configuration                                \\
		\hline
		Input                          &  1  &                                              \\
		\hline
		LowLevel                       & 128  & CONV 3$ \times $3, padding=1, PReLU             \\
		\hline
		\multirow{15}{*}{\shortstack{Enhanced\\DenseBlock-A}} & 16   & DCONV 3$ \times $3, dilation=1, padding=1, PReLU \\
		\cline{2-3}
		& 16   & DCONV 3$ \times $3, dilation=2, padding=2, PReLU \\
		\cline{2-3}
		& 32   & Concat, PReLU                                 \\
		\cline{2-3}
		& 16   & DCONV 3$ \times $3, dilation=3, padding=3, PReLU \\
		\cline{2-3}
		& 48   & Concat, PReLU                                 \\
		\cline{2-3}
		& 16   & DCONV 3$ \times $3, dilation=4, padding=4, PReLU \\
		\cline{2-3}
		& 64   & Concat, PReLU                                 \\
		\cline{2-3}
		& 16   & DCONV 3$ \times $3, dilation=4, padding=4, PReLU \\
		\cline{2-3}
		& 80   & Concat, PReLU                                 \\
		\cline{2-3}
		& 16   & DCONV 3$ \times $3, dilation=3, padding=3, PReLU \\
		\cline{2-3}
		& 96   & Concat, PReLU                                 \\
		\cline{2-3}
		& 16   & DCONV 3$ \times $3, dilation=2, padding=2, PReLU \\
		\cline{2-3}
		& 112  & Concat, PReLU                                 \\
		\cline{2-3}
		& 16   & DCONV 3$ \times $3, dilation=1, padding=1, PReLU \\
		\cline{2-3}
		& 128  & Concat, PReLU                                 \\
		\hline
		Concat-A                       & 256  &  LowLevel and enhanced DenseBlock-A                                         \\
		\hline
		Enhanced DenseBlock-B                   & 128  & Same as enhanced DenseBlock-A                        \\
		\hline
		Concat-B                       & 384  &   Concat-A and Enhanced DenseBlock-B                                          \\
		\hline
		Enhanced DenseBlock-C                   & 128  & Same as enhanced DenseBlock-A                              \\
		\hline
		Concat-C                       & 512  &  Concat-B and enhanced DenseBlock-C                                            \\
		\hline
		Bottleneck                     & 256  & CONV 1$ \times $1, padding=0                     \\
		\hline
		Reconstruction                 & 1    & CONV 3$ \times $3, padding=1                     \\                    
		\toprule[2pt]                   
	\end{tabular}
\end{table}

The overall architecture of the BDSS framework is displayed in Fig.~\ref{fig.BDSS framework}. The detailed configuration of the proposed model is provided in Table~\ref{Configuration}. BDSS can be decomposed into several parts: the convolution layer for extracting low-level features, three dense blocks for extracting high-level features, the bottleneck and reconstruction layers for generating the output. In each dense block, each dilated convolution layer is followed by a ParametricReLU (PReLU) \cite{he_delving_2015} as activation function except the bottleneck and reconstruction layers. The details of the proposed network structure is described in the following.

\subsubsection{Enhanced Dense Block}
Inspired by DenseNet structure \cite{huang_densely_2017}, we proposed a kind of enhanced dense block. The structure of enhanced dense block is displayed in Fig.~\ref{fig.Structure_2}. After adopting a convolution layer to the input speckled images for learning low-level features, three blocks are applied to extract the high-level features. Differing from ResNets \cite{he_deep_2016}, the feature maps extracted by different convolution layers are concatenated rather than directly summed.  In each dense block, the ${i^{th}}$ convolution layer (dilated convolution, described in the Section~\ref{Dilated}) receives the feature maps of all preceding layers as input:
\begin{equation}
{X_i} = {H_i}\left( {\left[ {{X_0},{X_1}, \ldots {X_{i - 1}}} \right]} \right),
\end{equation}
where $\left[ {{X_0},{X_1}, \ldots {X_{i - 1}}} \right]$ refers to the concatenation of the feature-maps produced in layers $0,1, \ldots i - 1$. ${H_i}\left(  \cdot  \right)$ denotes the composite function. 

In \cite{huang_densely_2017}, the composite function is decomposed into three layers: batch normalization (BN) \cite{ioffe_batch_nodate}, followed by a rectified linear unit (ReLU) and a $3 \times 3$ convolution (CONV). BN normalize the features using mean and variance in a batch during training and use estimated mean and variance of the whole training dataset during the testing. However, applying BN in image-to-image tasks is not optimal. BN tend to introduce unpleasant artifacts and limit the generation ability, which has proven in low-level computer vision problems such as PSNR-oriented super-resolution and deblurring tasks \cite{lim_enhanced_2017} \cite{nah_deep_2017}. Therefore, we remove BN to improve despeckling ability and reduce computational complexity and memory usage. In addition, we replace ReLU with PReLU and removed convolution in composite function. 

In the structure of the proposed model, short connections are added between a layer and every other layer. This kind of connectivity strengthens the flow of information through deep networks, thus alleviating the vanishing-gradient problem. In addition, the reuse of feature substantially reduced the number of parameters, therefore, we can reduce memory usage and computation as much as possible while increasing network depth for high performance.

\subsubsection{Dilated Convolution}
\label{Dilated}

\begin{figure}[] 
	\centering
	\includegraphics{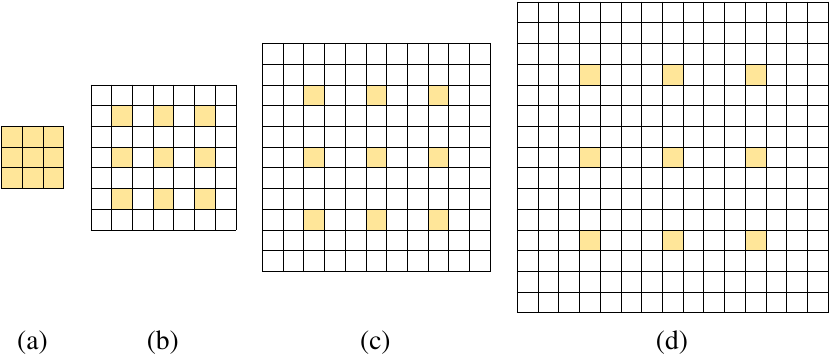} 
	\caption{Illustration of the dilated convolution. (a) corresponds to 1-dilated convolution, i.e., the common convolution; (b) corresponds to 2-dilated convolution; (c) corresponds to 3-dilated convolution; and (d) corresponds to 4-dilated convolution. The kernel sizes of their convolution are all $3 \times 3$. Obviously, the FOV becomes significantly larger as the dilation factor increases.}
	\label{fig.DCONV}
\end{figure}
For SAR despeckling, the context information can facilitate the reconstruction of the corrupted pixels. In CNN, enlarging the field of view (FOV) is the main way to augment the context information. 
Generically, there are two ways to achieve this purpose. We can either increase the size or depth of convolution. However, enlarging the size or increasing the depth both leads to more network parameters, which in turn increases computational complexity and memory usage. Thus, we introduce the dilated convolution (DCONV) \cite{yu_multi-scale_2015} to our model. The main idea of DCONV is to insert ``holes" (zeros) between filter elements, thereby, DCONV can increase the network's FOV while keeping the merits of convolution. 
Let $I$ be an input discrete 2-dimensional matrix, i.e., a speckled image. Let $k$ be a discrete filter of size ${{r}^2}$. The discrete convolution operator $ * $ can be defined as:
\begin{equation}
\left( {I*k} \right)\left( p \right) = \sum\nolimits_{s + t = p} {I\left( s \right)} k\left( t \right).
\end{equation}

Now, we replace CONV with DCONV, the $l$-dilated convolution operator ${{ * _l}}$ can be defined as:
\begin{equation}
\left( {I{*_l}k} \right)\left( p \right) = \sum\nolimits_{s + lt = p} {I\left( s \right)} k\left( t \right),
\end{equation}
where $l$ is a dilation factor.

For common convolution, its field of view $FO{V_c}$ is the same as the size of filter:
\begin{equation}
FO{V_c} = {r^2},
\end{equation}
and for dilated convolution, its field of view $FO{V_d}$ is:
\begin{equation}
FO{V_d} = {\left( {\left( {r + 1} \right)l - 1} \right)^2}.
\end{equation}

Fig.~\ref{fig.DCONV} gives the four kinds of $3 \times 3$ DCONV used in BDSS, their dilation factors are respectively set to 1, 2, 3, 4. For the 1-dilated convolution in Fig.~\ref{fig.DCONV} (a), i.e., the common convolution, its FOV is $3 \times 3$. And for others dilated convolutions in Fig.~\ref{fig.DCONV} (b), (c), (d), their FOVs are respectively $7 \times 7$, $11 \times 11$, and $15 \times 15$.

\begin{figure*}[tb] 
	\centering
	\includegraphics{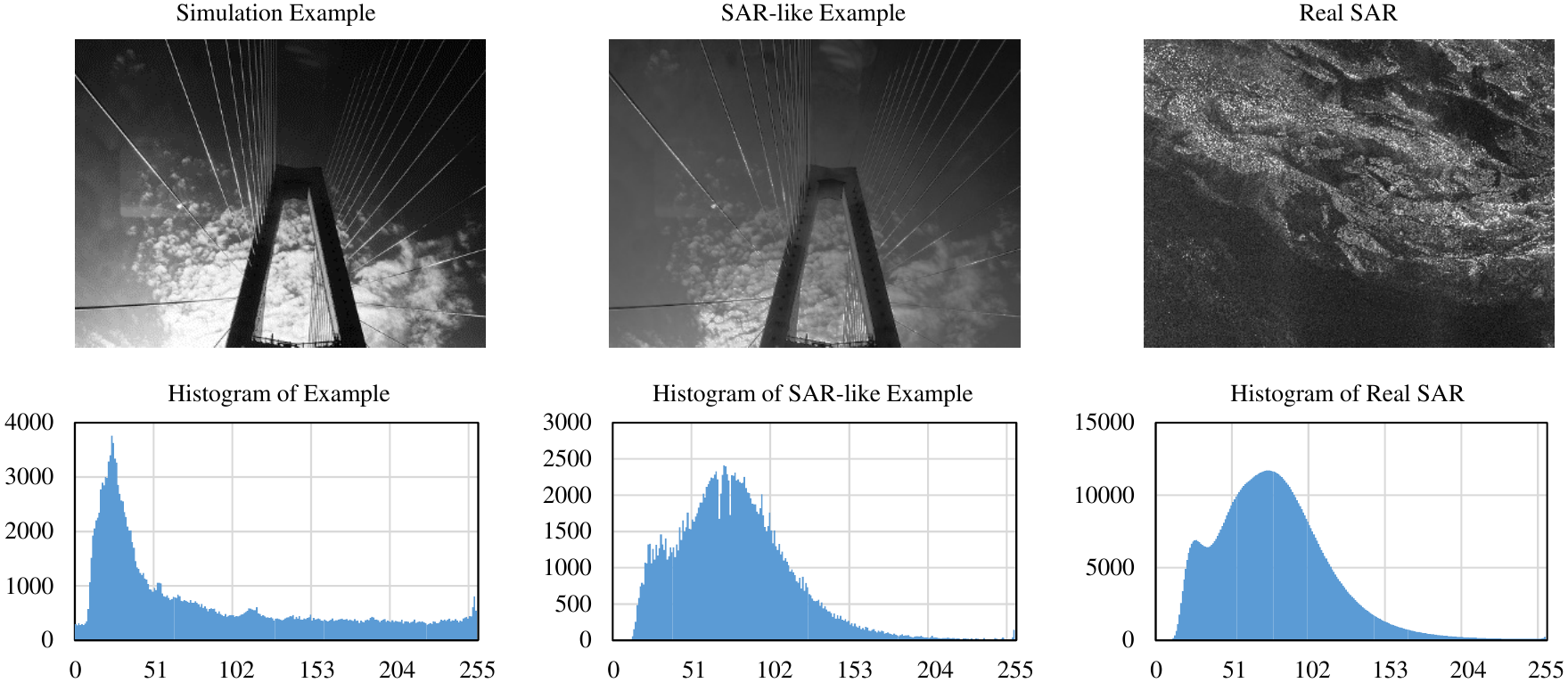} 
	\caption{Examples of simulated SAR-like images and SAR images.}
	\label{fig.SAR-like}
\end{figure*}

\section{Experiments and Analysis}
To verify the effectiveness of the proposed method, both synthetic and real-data experiments are performed, as described below. 
\subsection{Setup}
\subsubsection{SAR-like dataset}
The training data is 0.21 million images from the ILSVRC 2017 ImageNet \cite{ILSVRC15}. Regarding these images, we mainly refer to histograms of SAR images, transforming the ImageNet images into the SAR-like images with a similar distribution intensity as the training dataset. As shown in Fig.~\ref{fig.SAR-like}, whether the human visual observation or the histogram features of the corresponding images, it can be seen that the transformed images and the SAR images are extremely similar. For achieve blind despeckling in our proposed BDSS, the different speckle noise levels of looks $L = rand\left[ {1, + \infty } \right)$ are set up for training data. All images sent to the network for training are cropped to $112 \times 112$ pixels.

\subsubsection{Parameters Setting and Network Training}
The proposed model is trained using the Adam \cite{kingma_adam_2014} algorithm as the gradient descent optimization method, with momentum ${\beta _1} = 0.9$, ${\beta _2} = 0.999$, and $\varepsilon  = {10^{ - 8}}$. The learning rate $\alpha $ is initialized to 0.001 for the whole network and is decayed to half every three epochs. The training process of BDSS take 16 epochs. An epoch is equal to about 1.3\e{4} iterations, batch size is 16. We employ the PyTorch \cite{paszke2017automatic} framework to train the proposed BDSS on a PC with 128-GB RAM, an Intel Core i7-6900K CPU, and an NVIDIA 1080Ti GPU. 
\subsubsection{Assessment Indexes}
For the experiment on synthetic speckled images, the clean reference is $x$ available, so the performance assessment becomes much simpler. We choose two full-referenced indexes, the peak signal to noise ratio (PSNR) and structural similarity index (SSIM) \cite{wang_image_2004}. For the index of PSNR, a higher value means a better recovery of underlying signal, while a higher value of SSIM means a better recovery of underlying structural information.

For the experiment on real SAR images, no clean reference can be used, so the performance assessment only relies on some no-reference measures. In this paper, we evaluate performance of despeckling methods from four cases \cite{di_martino_benchmarking_2014} \cite{ma_review_2018}: 

Case 1: Speckle Reduction. The equivalent number of looks (ENL) \cite{oliver2004understanding} usually is used to assess the amount of speckle in SAR images, it is generally computed as:
\begin{equation}
ENL = {{{\mu ^2}} \over {{\sigma ^2}}},
\end{equation}
where $\mu $ and $\sigma $ respectively represent the mean and standard deviation of a homogeneous area. The larger value of the ENL, the better performance of speckle suppression.

Case 2: Point Target Feature Preservation. A strong point target is usually characterized by a cluster of pixels whose reflectivity values are much higher than the mean reflectivity of the surrounding scene. Based on this condition, a target-to-clutter ratio (TCR) is employed in \cite{argenti_tutorial_2013}, which measures the difference in the intensity ratios between point targets and the surrounding areas before and after despeckling by:
\begin{equation}
TCR=\left|20 \log _{10} \frac{\max _{p}\left(I_{d}\right)}{\operatorname{mean}_{p}\left(I_{d}\right)}-20 \log _{10} \frac{\max _{p}\left(I_{s}\right)}{\operatorname{mean}_{p}\left(I_{s}\right)}\right|,
\end{equation}
where ${I_s}$ and ${I_d}$ are respectively the speckled images and the despeckled images. Subscript $p$ denotes the patch containing a point target, and ${{{\max }_p}}$ and ${{{{\mathop{\rm mean}\nolimits} }_p}}$ are computed over the patch.

\begin{table*}[tb]	
	\renewcommand{\arraystretch}{1.3}
	\caption{Numerical Indexes for Synthetic Speckled Images}
	\centering
	\begin{tabular}{cccccccccc}		
		\toprule[2pt]		
		Looks&Index&Speckled Images&PPB-nonit&PPB-it25&SAR-BM3D&FANS&SAR-IDN&SAR-DRN&BDSS\\
		\hline
		\multirow{2}{*}{L=1}&PSNR&16.10&18.97&18.99&21.07&19.85&20.31&\color{blue}27.91&\color{red} 28.45\\
		\cline{2-10}
		&SSIM&0.3584&0.3556&0.3998&0.5332&0.5508&0.4972&\color{blue}0.8103&\color{red}0.8260\\
		\hline
		\multirow{2}{*}{L=2}&PSNR&18.76&20.99&21.17&23.52&22.79&22.29&\color{blue}29.27&\color{red}29.69\\
		\cline{2-10}
		&SSIM&0.4715&0.4653&0.5056&0.6391&0.6202&0.5942&\color{blue}0.8431&\color{red}0.8545\\
		\hline
		\multirow{2}{*}{L=4}&PSNR&21.52&23.21&23.55&25.80&24.98&24.46&\color{blue}30.55&\color{red}30.94\\
		\cline{2-10}
		&SSIM&0.5802&0.5692&0.6367&0.7169&0.6796&0.6966&\color{blue}0.8719&\color{red}0.8811\\
		\hline
		\multirow{2}{*}{L=8}&PSNR&24.33&25.40&25.87&27.80&26.96&26.23&\color{blue}31.90&\color{red}32.32\\
		\cline{2-10}
		&SSIM&0.6789&0.6630&0.7020&0.7753&0.7427&0.7793&\color{blue}0.8991&\color{red}0.9060\\		
		\toprule[2pt]				
	\end{tabular}
	\label{Tabel:Simulated}
\end{table*}

Case 3: Edge Feature Preservation. The edge-preservation degree based on the ratio of average (EPD-ROA) is given by \cite{xiangli_nie_variational_2015}:
\begin{equation}
EPD - ROA = {{\sum\limits_{i \in I} {\left| {{{{E_{DH}}\left( i \right)} \mathord{\left/
						{\vphantom {{{E_{DH}}\left( i \right)} {{E_{DV}}\left( i \right)}}} \right.
						\kern-\nulldelimiterspace} {{E_{DV}}\left( i \right)}}} \right|} } \over {\sum\limits_{i \in I} {\left| {{{{E_{SH}}\left( i \right)} \mathord{\left/
						{\vphantom {{{E_{SH}}\left( i \right)} {{E_{SV}}\left( i \right)}}} \right.
						\kern-\nulldelimiterspace} {{E_{SV}}\left( i \right)}}} \right|} }},
\end{equation}
where $I$ is the index set of a grey image, ${E_{DH}}\left( i \right)$ and ${E_{DV}}\left( i \right)$ represent the adjacent pixel values of the despeckled image along the horizontal and vertical direction, respectively. Similarly, ${E_{SH}}\left( i \right)$ and ${E_{SV}}\left( i \right)$ represent the corresponding adjacent pixel values of speckled images. The closer the value of EPD-ROA to 1, better the ability of edge preservation.

Case 4: Radiometric Preservation. A successful despeckling method should not significantly change the mean within a homogeneous region. A typical index for measuring the radiometric preservation capability is the mean of ratio (MOR). If the MOR value between the speckled image and the despeckled image significantly different from 1 indicates some radiometric distortion.

\subsubsection{Compared Methods}

Six despeckling methods are used for comparison. They are non-iterated and 25-iterated PPB filters \cite{deledalle_iterative_2009}, SAR-BM3D \cite{parrilli_nonlocal_2012}, fast adaptive nonlocal SAR despeckling (FANS) \cite{cozzolino_fast_2014}, SAR-IDN \cite{wang_sar_2017} and SAR-DRN \cite{zhang_learning_2018}. The first four methods can only achieve despeckling of SAR image with a specified look, and the other two CNN-based methods, SAR-IDN and SAR-DRN, can achieve blind despeckling of SAR image. In these methods, SAR-IDN and SAR-DRN are considered as the state of art. In following experiments, the parameter settings used in PPB filter, SAR-BM3D and FANS are set, respectively, as suggested in \cite{deledalle_iterative_2009}, \cite{parrilli_nonlocal_2012} and \cite{cozzolino_fast_2014}. For SAR-IDN and SAR-DRN, the learning rate, the number of training iterations and batch size are set to be the same as proposed BDSS, and other parameters settings are set as suggested in \cite{wang_sar_2017} and \cite{zhang_learning_2018}. The input training data of SAR-IDN and SAR-DRN is the same as proposed BDSS, and their ground truth data is original clean images without speckle noise.  

\subsection{Experiment on Synthetic Speckled Images}
\label{Synthetic}

\begin{figure*}[htb] 
	\centering 
	\includegraphics{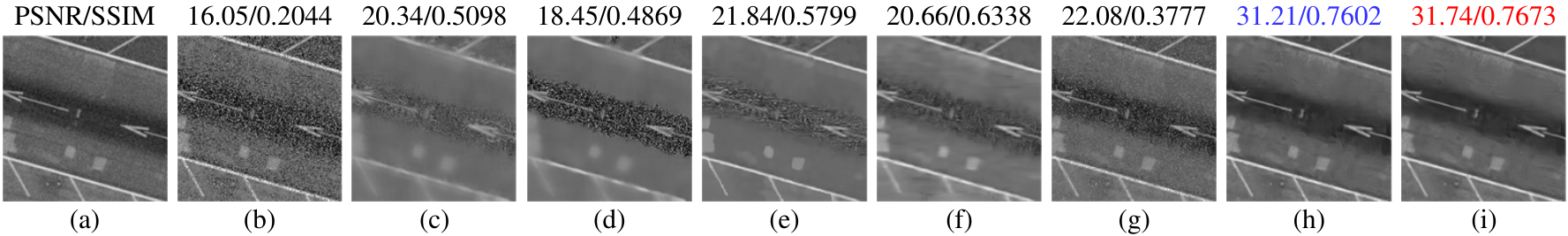} 
	\caption{Results for the Runway image contaminated by 1-look speckle. (a) Original clean image. (b) Speckled image. (c) PPB-nonit. (d) PPB-it25. (e) SAR-BM3D. (f) FANS. (g) SAR-IDN. (h) SAR-DRN. (i) BDSS.}
	\label{Runway}
\end{figure*}

\begin{figure*}[htb] 
	\centering 
	\includegraphics{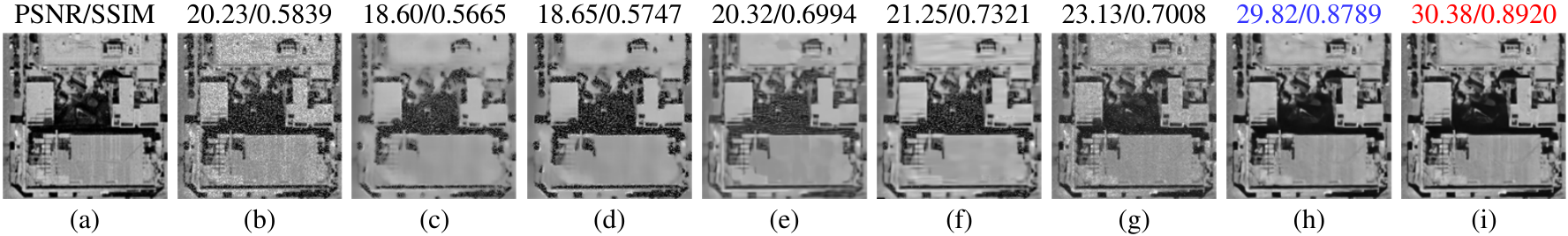} 
	\caption{Results for the Building image contaminated by 2-look speckle. (a) Original clean image. (b) Speckled image. (c) PPB-nonit. (d) PPB-it25. (e) SAR-BM3D. (f) FANS. (g) SAR-IDN. (h) SAR-DRN. (i) BDSS.}
	\label{Building}
\end{figure*}

\begin{figure*}[htb] 
	\centering 
	\includegraphics{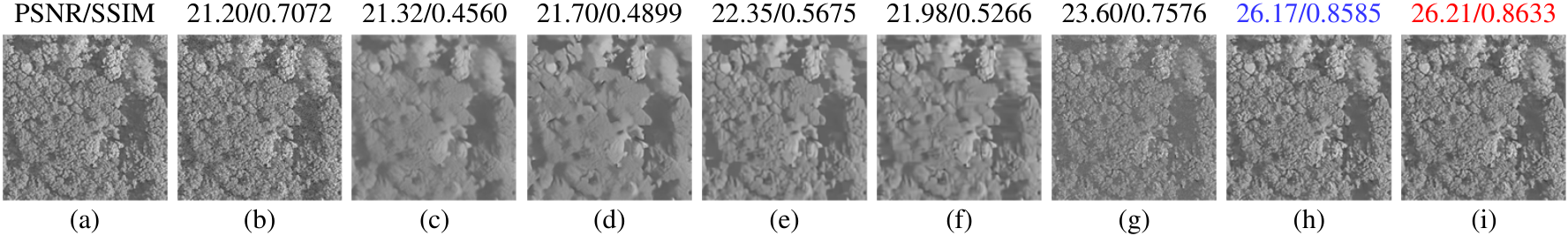} 
	\caption{Results for the Forest image contaminated by 4-look speckle. (a) Original clean image. (b) Speckled image. (c) PPB-nonit. (d) PPB-it25. (e) SAR-BM3D. (f) FANS. (g) SAR-IDN. (h) SAR-DRN. (i) BDSS.}
	\label{Forest}
\end{figure*}

\begin{figure*}[htb] 
	\centering 
	\includegraphics{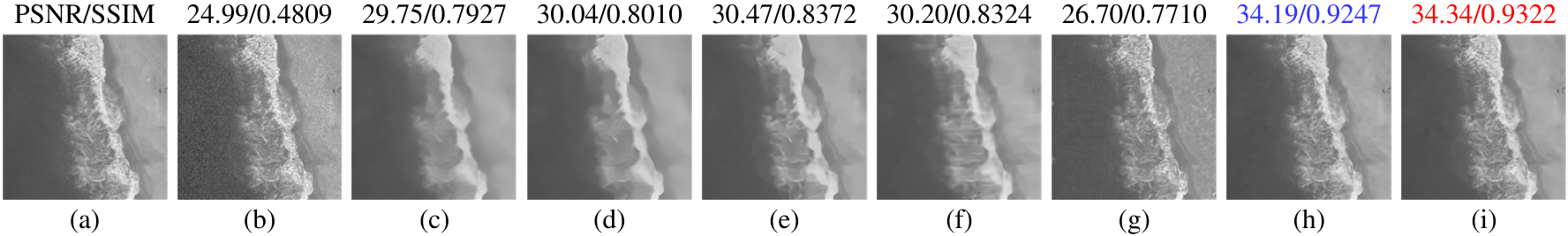} 
	\caption{Results for the Beach image contaminated by 8-look speckle. (a) Original clean image. (b) Speckled image. (c) PPB-nonit. (d) PPB-it25. (e) SAR-BM3D. (f) FANS. (g) SAR-IDN. (h) SAR-DRN. (i) BDSS.}
	\label{Beach}
\end{figure*}

The use of synthetic speckled images allows an objective performance assessment of the speckle suppression efficiency. To achieve this goal, the common approach in the literature is to use a virtually noiseless optical image as a clean reference and to inject speckle on it. In this experiment, we use UC Merced land use dataset \cite{yang_bag--visual-words_2010} for the test, which of images are manually extracted from large images from the USGS National Map Urban Area Imagery collection for various urban areas around the country, and which of the pixel resolution is 1 foot. The dataset has 21 classes land -use image and each class has 100 images with a size of $256 \times 256$ pixels. We choose 360 images (18 classes, each class have 20 images) for the test. To verify the despeckling effectiveness with known speckle look, we set four different speckle noise levels of looks $L = 1,2,4$ and $8$. To acquire an integrated comparison for the other methods and the proposed BDSS, quantitative evaluation indexes (PSNR and SSIM) and a visual comparison are used to analyze the results of different methods. In Table~\ref{Tabel:Simulated}, we present the averages of contrasting evaluation indexes of the four different speckle noise level. To give detailed contrasting results, a Runway image with $L = 1$, a Building image with $L = 2$, a Forest image with $L = 4$ and a Beach image with $L = 8$ are chosen to demonstrate the visual results, corresponding to Figs.~\ref{Runway}, \ref{Building}, \ref{Forest} and \ref{Beach}. In Table~\ref{Tabel:Simulated}, the best performance for each quality index is marked in red and the second-best performance for each quality index is blue. 

As shown in Table~\ref{Tabel:Simulated}, the proposed BDSS obtains all the best PSNR and SSIM results in four speckle noise levels, and SAR-DRN obtains second-best results. The PSNR and SSIM results of these methods have all improved with the increase of speckle noise levels. For PPB filters, whether noniterated and 25-iterated PPB filter, their PSNR results have a certain degree of improvement compared with input speckled images. However, their SSIM results have not been significantly improved, even reduced. Specifically, the SSIM results of PPB-nonit underperforms input speckled images by about 0.01, and these of PPB-it25 outperforms only input speckled images by about 0.04. SAR-BM3D, FANS, and SAR-IDN perform moderately, of which of SAR-BM3D is a better method. Compared with SAR-BM3D, the performance of supervised SAR-DRN and our proposed method BDSS has been greatly improved, by at least 5 and 0.1 on PSNR and SSIM, respectively. In addition, the difference in speckle noise level has less impact on BDSS and SAR-DRN. Compared with $L=8$, the results of BDSS on PSNR and SSIM are reduced by about 0.4 and 0.08 respectively when $L=1$, while the results of SAR-BM3D are reduced by about 6.8 and 0.24 respectively. Though never seen clean reference, our proposed BDSS even perform better, compared with state-of-the-art SAR-DRN. 

Figs.~\ref{Runway}, \ref{Building}, \ref{Forest} and \ref{Beach} give some visual samples. Fig.~\ref{Runway} (a) is a Runway image contaminated by 1-look speckle. This image contains some ground signs with distinct edge feature such as an arrow. Two PPB filters, SAR-BM3D, and FANS simultaneously result in a certain degree of distortion. SAR-IDN performs well on edge feature preservation, while does not effectively despeckling. As shown in Fig.~\ref{Runway} (h) and (i), SAR-DRN and BDSS can keep ground signs clear while despeckling effectively.  A similar situation is also reflected in Fig.~\ref{Building}, which is more obvious. The PSNR and SSIM values of two PPB filters are lower with the input speckled images. The SSIM values of SAR-BM3D and FANS are improved obviously, while PSNR values are not. Fig.~\ref{Forest} (a) is a Forest image contaminated by 4-look speckle, which contains lots of texture detail. As shown in Fig.~\ref{Forest} (c), (d), (e), and (f), two PPB filters, SAR-BM3D, and FANS loss much texture detail in order to image smoothing. This is reflected in that, their PSNR values are improved compared with Fig.~\ref{Forest} (b) (input speckled images), while their SSIM values are reduced. Fig.~\ref{Beach} (a) is a Beach image contaminated by 8-look speckle, which contains a homogeneous sea area, a homogeneous land area and a wave area with much texture detail. Similarly, we can see that two PPB filters, SAR-BM3D, and FANS remove much speckle noise in two homogeneous, while loss much texture detail in wave area. SAR-IDN has the opposite performance. SAR-DRN and proposed BDSS perform well on despeckling and texture detail preservation, and BDSS performs better.

\subsection{Experiment on Real SAR Images}

\begin{figure*}[htbp] 
	\centering 
	\includegraphics{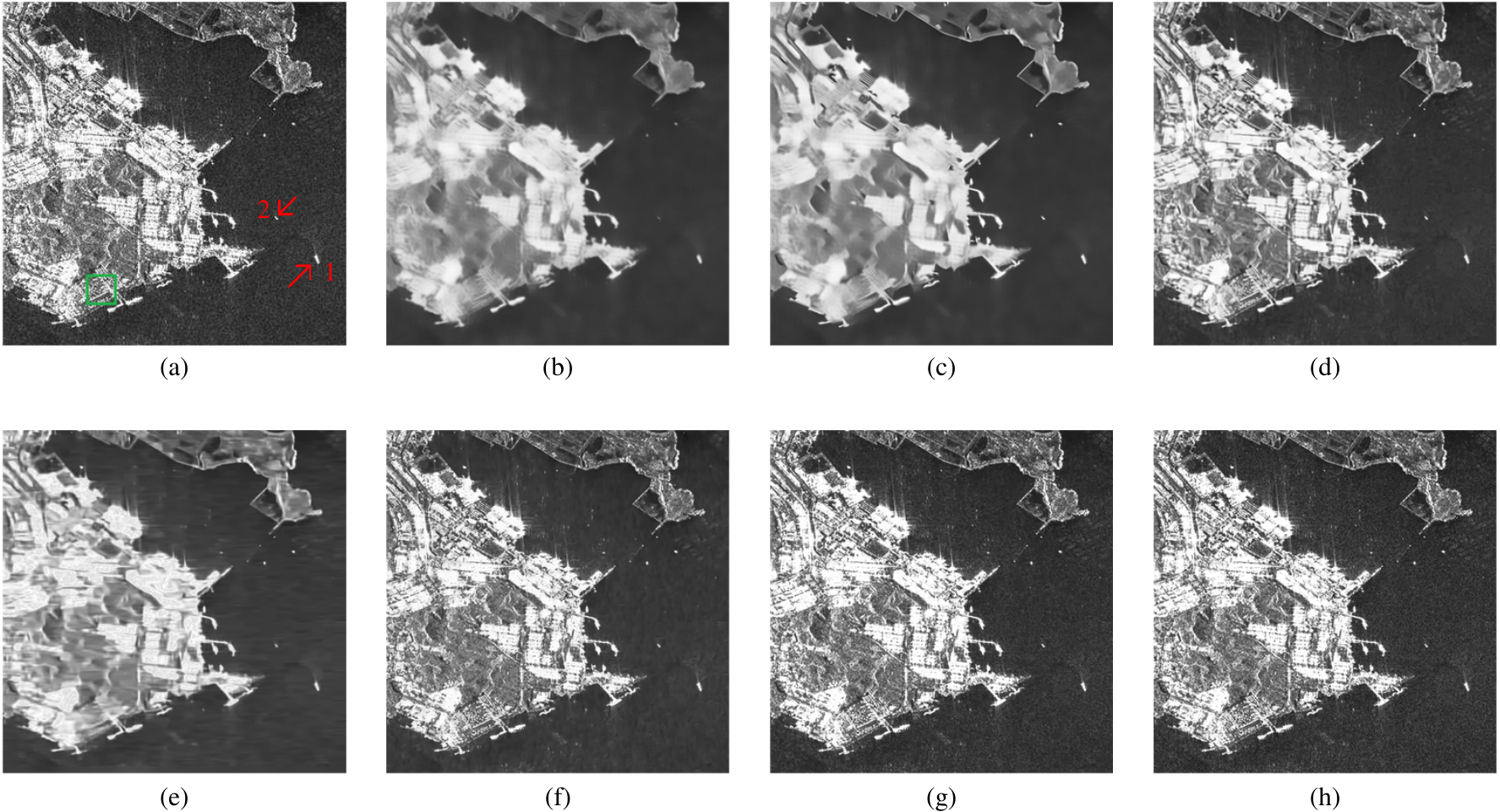} 
	\caption{Results for the Sentinel-1 image contaminated by 1-look speckle. (a) Original speckled image. (b) PPB-nonit. (c) PPB-it25. (d) SAR-BM3D. (e) FANS. (f) SAR-IDN. (g) SAR-DRN. (h) BDSS.}
	\label{Sentinel-1}
\end{figure*}

\begin{figure*}[htbp] 
	\centering 
	\includegraphics{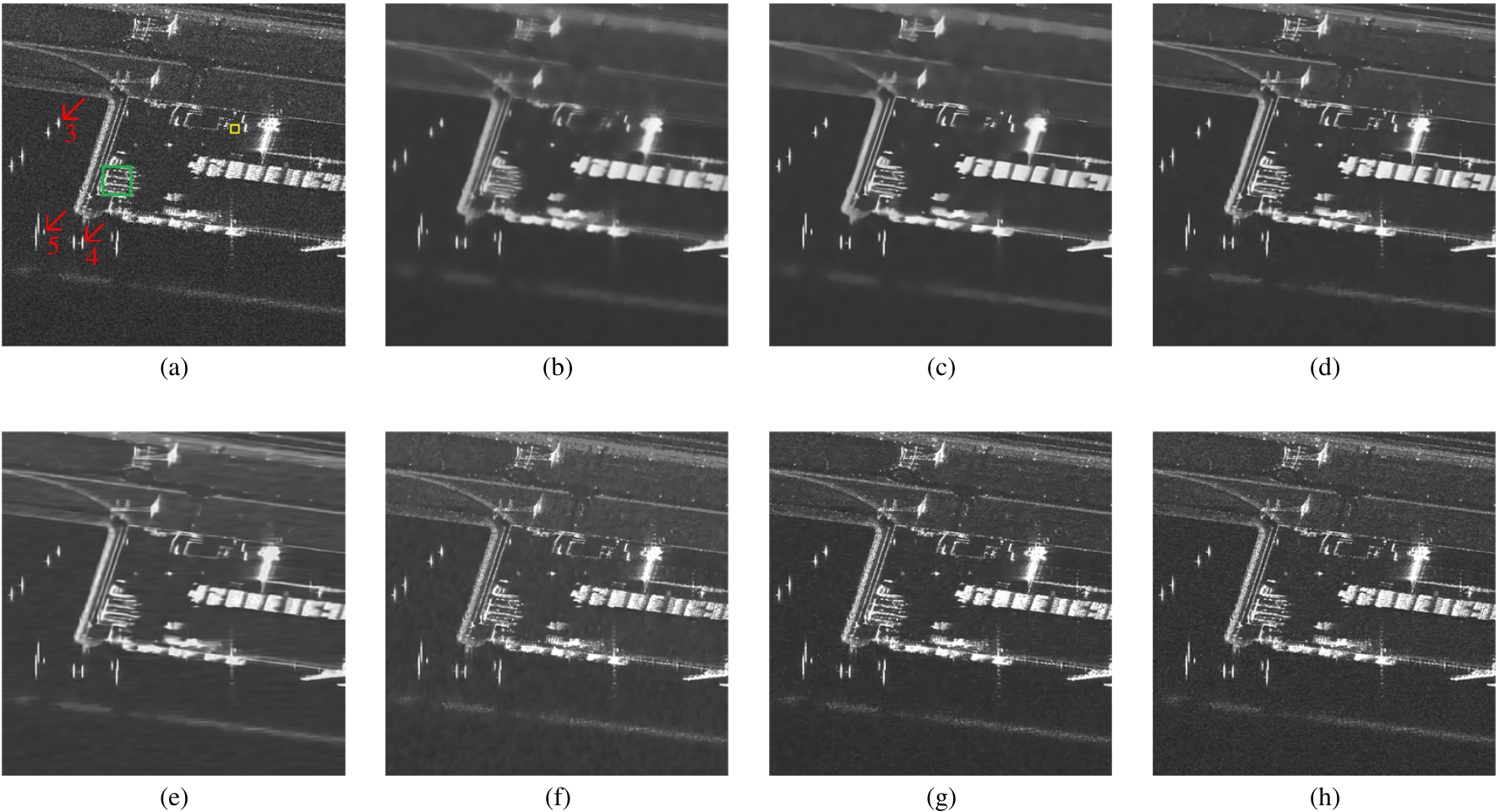} 
	\caption{Results for the TerraSAR image contaminated by 1-look speckle. (a) Original speckled image. (b) PPB-nonit. (c) PPB-it25. (d) SAR-BM3D. (e) FANS. (f) SAR-IDN. (g) SAR-DRN. (h) BDSS.}
	\label{TerraSAR-X}
		
\end{figure*}

\begin{figure*}[htbp] 
	\centering 
	\includegraphics{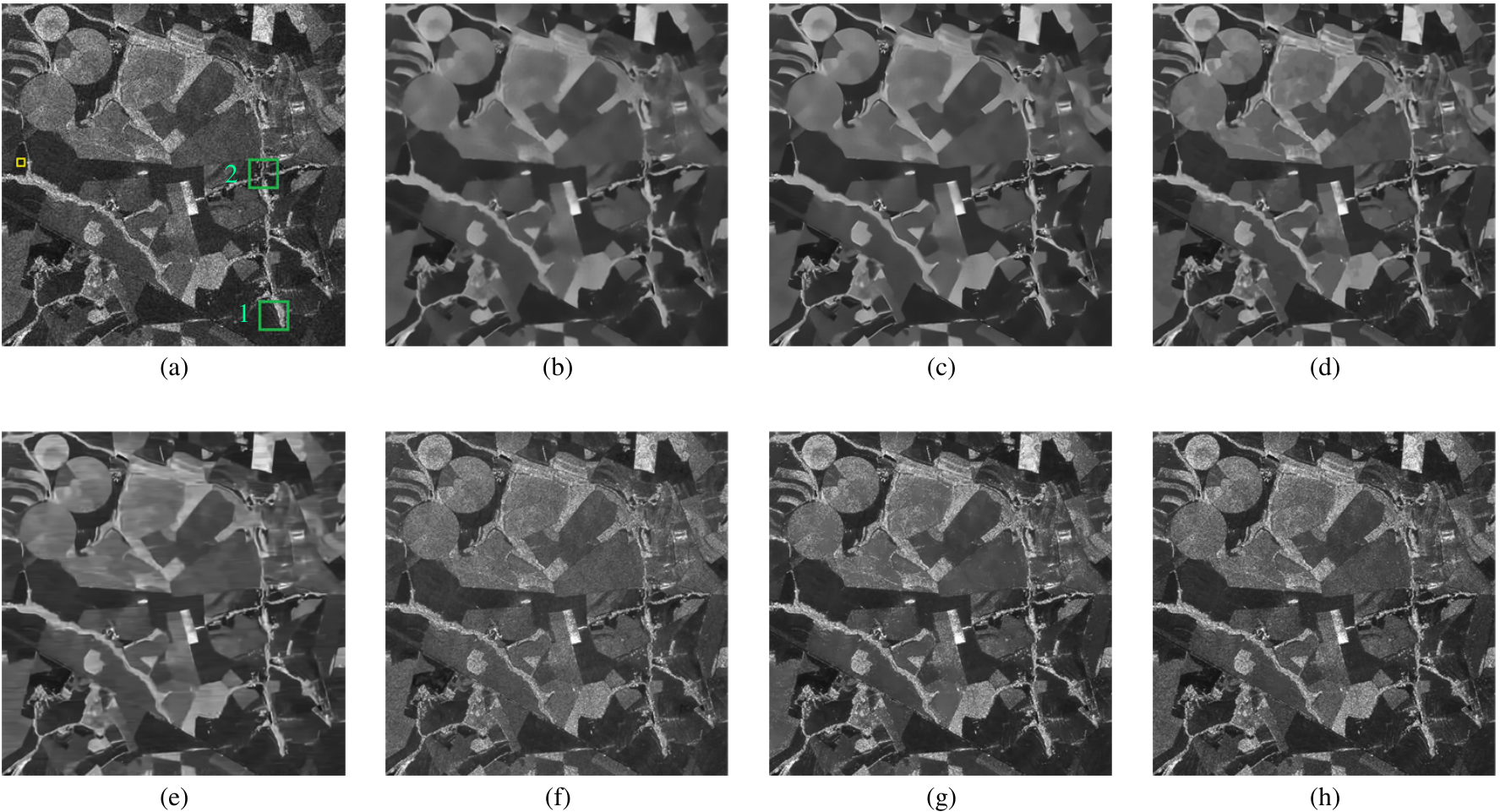} 
	\caption{Results for the ALOS-2 image contaminated by 2-look speckle. (a) Original speckled image. (b) PPB-nonit. (c) PPB-it25. (d) SAR-BM3D. (e) FANS. (f) SAR-IDN. (g) SAR-DRN. (h) BDSS.}
	\label{ALOS}

\end{figure*}

\begin{figure*}[htbp] 
	\centering 
	\includegraphics{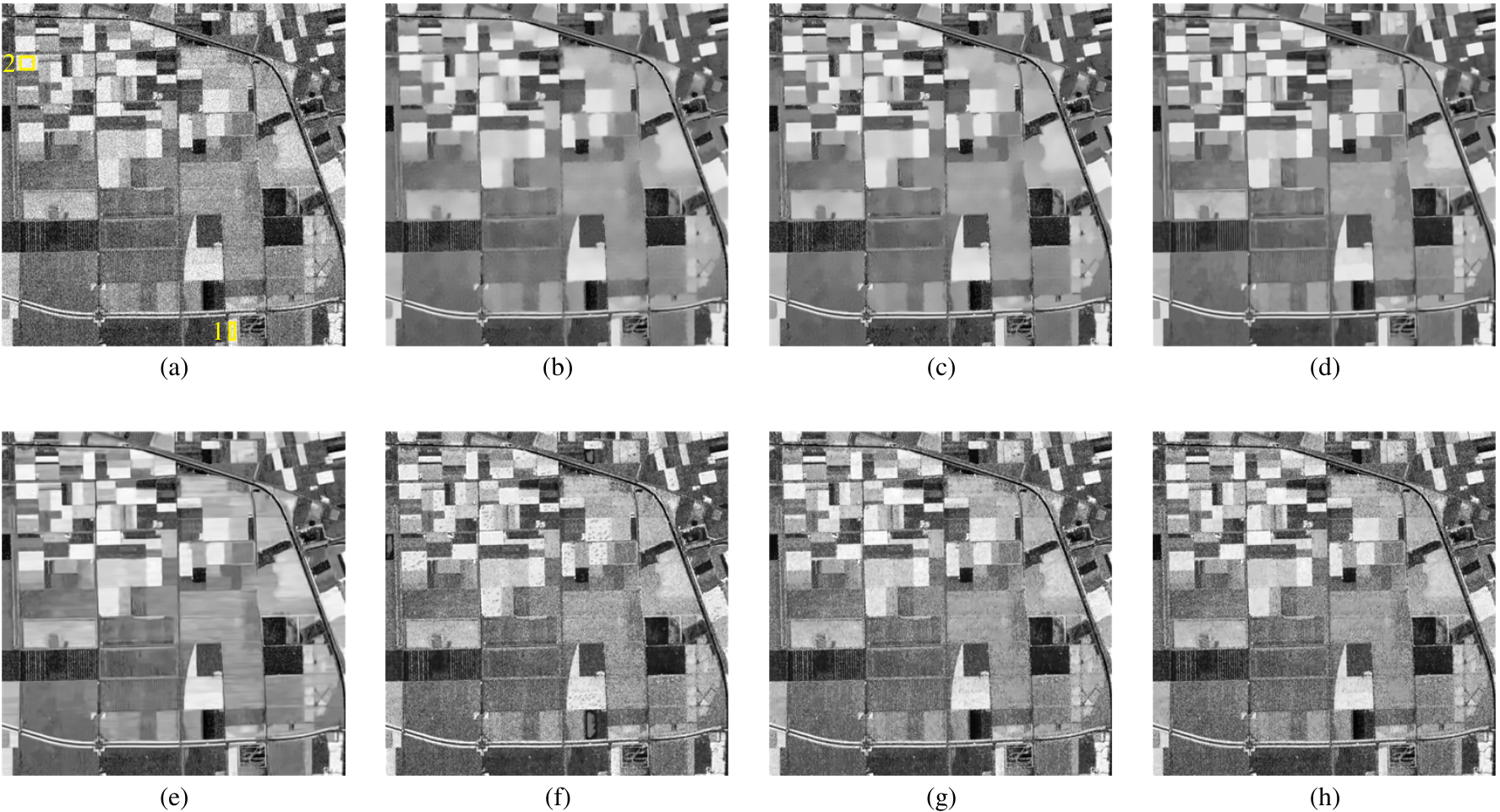} 
	\caption{Results for the AIRSAR image contaminated by 4-look speckle. (a) Original speckled image. (b) PPB-nonit. (c) PPB-it25. (d) SAR-BM3D. (e) FANS. (f) SAR-IDN. (g) SAR-DRN. (h) BDSS.}
	\label{AIRSAR}

\end{figure*}

\begin{table*}[htbp]
	\renewcommand{\arraystretch}{1.3}
	\caption{ENL Indexes for Real SAR Images}
	\centering
	\begin{tabular}{cccccccccc}
		\toprule[2pt]
		\multicolumn{2}{c}{Data}             & Original & PPB-nonit & PPB-it25 & SAR-BM3D & FANS    & SAR-IDN & SAR-DRN & BDSS    \\ \hline 
		\multicolumn{2}{c}{TerraSAR}   & 16.79    & \color{blue}976.91      & \color{red}{1511.19}     & 899.04    & 441.58  & 157.49  & 268.95  & 295.31  \\ \hline 
		\multicolumn{2}{c}{ALOS-2}         & 22.57    & \color{blue}16440.13    & \color{red}{23932.63}    & 24055.05  & 959.28  & 382.66  & 193.43  & 583.69  \\ \hline 
		\multirow{2}{*}{AIRSAR} & Region 1 & 213.07   & \color{blue}{55156.8}     & 16652.38    & \color{red}{564467.56} & 6962.56 & 420.02  & 1355.79 & 2415.12 \\ \cline{2-10} 
		& Region 2 & 72.86    & \color{blue}{21312.28}    & 8468.95     & \color{red}{152451.39} & 4165.91 & 185.46  & 612.41  & 1863.71 \\ \hline 
		\multicolumn{2}{c}{Mean}        & 84.14    & \color{blue}{24191.28}    & 14032.07    & \color{red}{196473.88} & 2787.80 & 320.06  & 606.06  & 1098.04 \\ 
		\toprule[2pt]
				\label{Tabel:ENL}
	\end{tabular}
\end{table*}

\begin{table*}[htbp]
	\renewcommand{\arraystretch}{1.3}
	\caption{EPD-ROA Indexes for Real SAR Images}
	\centering
	\begin{tabular}{ccccccccc}
				\toprule[2pt]
		\multicolumn{2}{c}{Data}& PPB-nonit & PPB-it25 & SAR-BM3D & FANS   & SAR-IDN & SAR-DRN & BDSS   \\
				\hline
		\multicolumn{2}{c}{Sentinel-1} & 0.7915      & 0.7917      & 0.8060    & 0.7918 & 0.8658  & \color{blue}{0.8824}  & \color{red}{0.9145 }\\
				\hline
\multicolumn{2}{c}{TerraSAR} & 0.8337      & 0.8960      & 0.9016    & 0.8606 & 0.9188  & \color{blue}{0.9583}  & \color{red}{0.9803} \\
				\hline
		\multirow{2}{*}{ALOS-2}& Region 1    & 0.9191      & 0.9343      & 0.9288    & 0.9215 & 0.9348  & \color{blue}{0.9652}  & \color{red}{0.9774} \\
			 \cline{2-9} 
&Region 2    & 0.9137      & 0.9474      & 0.9288    & 0.9292 & 0.9343  & \color{blue}{0.9656}  & \color{red}{0.9774} \\
				\hline
\multicolumn{2}{c}{Mean}         & 0.8645      & 0.8923      & 0.8913    & 0.8758 & 0.9134  & \color{blue}{0.9429}  & \color{red}{0.9624}\\
				\toprule[2pt]
	\end{tabular}
	\label{Tabel:EPD-ROA}
\end{table*}

\begin{table*}[htbp]

	\renewcommand{\arraystretch}{1.3}
	\caption{TCR Indexes for Real SAR Images}
	\centering
	
	\begin{tabular}{ccccccccc}
		\toprule[2pt]
	\multicolumn{2}{c}{Data}& PPB-nonit                & PPB-it25                & SAR-BM3D                  & FANS                       & SAR-IDN                    & SAR-DRN                    & BDSS                       \\
			\hline
		\multirow{2}{*}{Sentinel-1} & Point Target 1                  & 0.5849                     & 0.2499                     & \color{blue}{0.0914}                     & 0.7039                     & 0.4603                     & 0.1434                     & \color{red}{0.0873}                     \\
				\cline{2-9}
		& Point Target 2                  & 5.4117                     & 0.4245                     & 0.2748                     & 2.3786                     & 0.6111                     & \color{blue}{0.1296 }                    & \color{red}{0.1052}                     \\
			\hline
		\multirow{3}{*}{TerraSAR} & Point Target 3                  & 0.6628                     & 0.1821                     & \color{blue}{0.1338}                     & 0.6509                     & 0.9046                     & 0.1870                     & \color{red}{0.0100}                     \\
			\cline{2-9}
		& Point Target 4                  & 0.6097                     & 0.0768                     & \color{blue}{0.1011}                     & 0.5657                     & 0.5850                     & 0.1857                     & \color{red}{0.0320}                     \\
				\cline{2-9}
		& Point Target 5                  & 2.1103                     & 0.3009                     & \color{blue}{0.1242}                     & 1.1786                     & 1.2492                     & 0.2788                     & \color{red}{0.0173}                     \\
			\hline
		\multicolumn{2}{c}{Mean}    & 1.3319      & 0.1713      & 0.1000    & 0.7467 & 0.3952  & \color{blue}{0.0920}  & \color{red}{0.0405}\\
		\toprule[2pt]
	\end{tabular}
	\label{Tabel:TCR}
\end{table*}

\begin{table*}[htbp]
	
	\renewcommand{\arraystretch}{1.3}
	\caption{MOR Indexes for Real SAR Images}
	\centering	
	\begin{tabular}{ccccccccc}
		\toprule[2pt]
	\multicolumn{2}{c}{Data}& PPB-nonit & PPB-it25 & SAR-BM3D & FANS   & SAR-IDN & SAR-DRN & BDSS   \\
		                	\hline
	\multicolumn{2}{c}{TerraSAR}  & 1.1706      & 1.1791      & 1.1134    & 1.2243 & 1.2198  & \color{blue}{1.0440}  & \color{red}{1.0238} \\
			\hline
	\multicolumn{2}{c}{ALOS-2}         & 1.1123      & 1.1069      & 1.0882    & 1.1409 & 1.0483  & \color{blue}{1.0307}  & \color{red}{1.0092} \\
			\hline
	\multirow{2}{*}{AIRSAR} & Region 1 & 1.1141      & 1.0838      & 1.1085    & 1.0973 & 1.0544  & \color{blue}{1.0457}  & \color{red}{1.0220} \\
		\cline{2-9} 
	& Region 2 & 1.0272      & 1.0301      & \color{blue}{1.0291}    & 1.0533 & 1.0450  & 1.0353  & \color{red}{1.0241} \\
			\hline
	\multicolumn{2}{c}{Mean}        & 1.1323      & 1.1233      & 1.1034    & 1.1542 & 1.1075  & \color{blue}{1.0401}  & \color{red}{1.0183}\\
		\toprule[2pt]
	\end{tabular}
\label{Tabel:MOR}
		
\end{table*}

In this section, to further verify the effectiveness of the proposed BDSS, four real SAR images obtained at different scenes with different remote sensors are used for evaluation. The real SAR images are: 1) a harbor nearby Lianyungang, China. (downloaded from https://sentinel.esa.int/web/sentinel/missions/sentinel-1/data-products, obtained by Sentinel-1, C-band, one-look, as shown in Fig.~\ref{Sentinel-1} (a)); 2) a harbor nearby Rotterdam, Netherlands. (downloaded from https://www.intelligence-airbusds.com, obtained by TerraSAR, X-band, one-look, as shown in Fig.~\ref{TerraSAR-X} (a)); 3) an agriculture area nearby Brazil. (downloaded from https://www.eorc.jaxa.jp, obtained by ALOS-2, L-band, 2-look, as shown in Fig.~\ref{ALOS} (a)); and 4) an agricultural area nearby Flevoland, Netherlands. (downloaded from https://earth.esa.int, obtained by AIRSAR, L-band, 4-look, as shown in Fig.~\ref{AIRSAR} (a)). These images are all cropped to $600\times600$ pixels. The corresponding visual results are given in Figs~\ref{Sentinel-1}, \ref{TerraSAR-X}, \ref{ALOS} and \ref{AIRSAR}.

Due to the lack of the true signal of real SAR images, the indexes including ENL, EPD-ROA, TCR, and MOR are adopted for evaluation. The numerical results are given in Tables~\ref{Tabel:ENL}, \ref{Tabel:EPD-ROA}, \ref{Tabel:TCR}, and \ref{Tabel:MOR}. Moreover, the ENL and MOR indexes are obtained by four homogeneous regions that are manually selected in Figs.~\ref{TerraSAR-X} (a), \ref{ALOS} (a) and \ref{AIRSAR} (a) (identified by yellow rectangles). The EPD-ROA indexes are obtained by four regions with much edge features in Figs.~\ref{Sentinel-1} (a), \ref{TerraSAR-X} (a), and \ref{ALOS} (a) (identified by green rectangles). The TCR indexes are obtained by five point targets on the sea in Figs.~\ref{Sentinel-1} (a) and \ref{TerraSAR-X} (a) (identified by red arrows). The best performance for each quality index is marked in red and the second-best performance for each quality index is blue. 

As shown in Table~\ref{Tabel:ENL}, the ENL values of two PPB filters, SAR-BM3D and FANS are quite huge which of the largest is even hundreds of thousands, while the ENL values of original input SAR images are only about 200 even lower. This phenomenon is also shown in Figs~\ref{TerraSAR-X}, \ref{ALOS} and \ref{AIRSAR} (b)-(e), it seems that over-smoothing exist for these methods. In the above experiment on synthetic speckled images, their lower SSIM values also confirm this. This kind of over-smoothing image is obviously not desirable in practical applications such as target detection and classification. An ideal method should try to despeckling while maintaining well feature such point target, edge and radiometric. 

SAR-IDN, SAR-DRN, and our proposed BDSS do well on the trade-off between despeckling and feature preservation. From the mean EPD-ROA, TCR and MOR values shown in Tables~\ref{Tabel:EPD-ROA}, \ref{Tabel:TCR} and \ref{Tabel:MOR}, BDSS perform best on edge feature, point target, and radiometric preservation. SAR-DRN perform second-best, and SAR-IDN performs third-best on edge and radiometric preservation, fourth-best on point target preservation. From the ENL values shown in Table~\ref{Tabel:ENL}, we can see that proposed BDSS perform best on speckle reduction compared with SAR-IDN and SAR-DRN.

Based on the analysis above, we can conclude that the proposed BDSS method can notably suppress speckle with well feature preservation such as edge, point target, and radiometric. SAR-DRN has comparable performance with BDSS.

\subsection{Blind Despeckling Performance Analysis}

To verify the blind despeckling effectiveness of the proposed method, we design an experiment. We add random speckle noise levels of looks $L = random\left[ {1,10} \right]$ on clean optical images, which are same as with these in Section~\ref{Synthetic}. Tabel~\ref{Blind} gives the PSNR and SSIM values between original clean images and despeckled images with different methods. With a certain SAR image, the known number of looks is necessary for PPB, SAR-BM3D, and FANS. Therefore, we set four looks $L = 1, 2, 4, 8$ for each method respectively. From the Table~\ref{Blind}, we can see that, as the number of looks set by these methods increase, the PSNR and SSIM values also increases. For PPB-nonit with $L=1, 2$, PPB-it25 with $L=1$, SAR-BM3D with $L=1$, FANS with $L=1, 2$ and SAR-IDN, their PSNR values are not improved compared with input speckled images. Similarly, for PPB-nonit with $L=1, 2, 4$, PPB-it25 with $L=1, 2$, SAR-BM3D with $L=1, 2$, FANS with $L=1, 2$ and SAR-IDN, their SSIM values are not improved. Compared with input speckled images, the PSNR and SSIM values of proposed BDSS has a significant improvement, up to 10.83 and 0.3799 respectively. SAR-DRN has a second-best performance. Though without clean references in the training process, our proposed BDSS has a better blind despeckling performance than SAR-DRN. The possible reason is that, in order to complete despeckling for different speckle noise levels, SAR-DRN based on supervised learning, needs to learn multiple mappings (different levels of speckle noise to clean). However, BDSS based self-supervised learning only need to learn a kind of mapping (speckle noise to speckle noise). 

\begin{table}[htbp]
	\renewcommand{\arraystretch}{1.3}
	\caption{Results of Blind Despeckling Test}
	\centering	
	\begin{tabular}{cccc}
				\toprule[2pt]
		\multicolumn{2}{c}{Indexes} & \multicolumn{1}{c}{PSNR} & \multicolumn{1}{c}{SSIM} \\\hline
		\multicolumn{2}{c}{Speckled images}                & 18.95                    & 0.4752                   \\\hline
		\multirow{4}{*}{PPB-nonit} & L=1                  & 17.29                    & 0.2175                   \\\cline{2-4}
		& L=2                  & 18.20                    & 0.3250                   \\\cline{2-4}
		& L=4                  & 19.10                    & 0.4297                   \\\cline{2-4}
		& L=8                  & 19.91                    & 0.5286                   \\\hline
		\multirow{4}{*}{PPB-it25}  & L=1                  & 19.13                    & 0.3459                   \\\cline{2-4}
		& L=2                  & 20.27                    & 0.4446                   \\\cline{2-4}
		& L=4                  & 21.36                    & 0.5422                   \\\cline{2-4}
		& L=8                  & 22.54                    & 0.6430                   \\\hline
		\multirow{4}{*}{SAR-BM3D}  & L=1                  & 18.80                    & 0.3760                   \\\cline{2-4}
		& L=2                  & 19.26                    & 0.4239                   \\\cline{2-4}
		& L=4                  & 20.14                    & 0.5053                   \\\cline{2-4}
		& L=8                  & 22.28                    & 0.6572                   \\\hline
		\multirow{4}{*}{FANS}      & L=1                  & 17.18                    & 0.3650                   \\\cline{2-4}
		& L=2                  & 18.88                    & 0.4444                   \\\cline{2-4}
		& L=4                  & 20.61                    & 0.5347                   \\\cline{2-4}
		& L=8                  & 23.20                    & 0.6636                   \\\hline
		\multicolumn{2}{c}{SAR-IDN}                       & 19.13                    & 0.3459                   \\\hline
		\multicolumn{2}{c}{SAR-DRN}                       &  \color{blue}29.35                    &  \color{blue}0.8439                   \\\hline
		\multicolumn{2}{c}{BDSS}                          &  \color{red}29.78                    &  \color{red}0.8551\\         
				\toprule[2pt]         
	\end{tabular}
\label{Blind}
\end{table}

\section{Conclusion}
In this paper, self-supervised learning is first introduced to achieve blind despeckling of SAR image. Differing from the other deep learning-based methods, inputs and references in BDSS's training process both are drawn from a distribution corrupted by speckle noise. Due to the multiplicative speckle noise in a SAR image satisfies unit mean distribution, BDSS can output the arithmetic mean of the speckled images, i.e., clean images. Dense connection is employed to reduce memory usage as much as possible while increasing network depth for high performance. Dilated convolutions are used to enlarge the receptive field. To train BDSS, a SAR-like dataset based ImageNet is created, which is similar to real SAR images in whether human visual observation or the statistical distribution. We design two experiments on synthetic speckled images and real SAR images. Comparing with some the-state-of-art despeckling methods, a reasonable performance is achieved by our proposed BDSS, in terms of the speckle reduction and the preservation of edges, point targets, and radiometric. In addition, we give the experiment results of random speckle level (look) with different methods, BDSS do best in blind despeckling.


%

%
%

\ifCLASSOPTIONcaptionsoff
  \newpage
\fi



\bibliographystyle{IEEEtran}
\bibliography{paper}

\begin{thebibliography}{10}
\providecommand{\url}[1]{#1}
\csname url@samestyle\endcsname
\providecommand{\newblock}{\relax}
\providecommand{\bibinfo}[2]{#2}
\providecommand{\BIBentrySTDinterwordspacing}{\spaceskip=0pt\relax}
\providecommand{\BIBentryALTinterwordstretchfactor}{4}
\providecommand{\BIBentryALTinterwordspacing}{\spaceskip=\fontdimen2\font plus
\BIBentryALTinterwordstretchfactor\fontdimen3\font minus
  \fontdimen4\font\relax}
\providecommand{\BIBforeignlanguage}[2]{{%
\expandafter\ifx\csname l@#1\endcsname\relax
\typeout{** WARNING: IEEEtran.bst: No hyphenation pattern has been}%
\typeout{** loaded for the language `#1'. Using the pattern for}%
\typeout{** the default language instead.}%
\else
\language=\csname l@#1\endcsname
\fi
#2}}
\providecommand{\BIBdecl}{\relax}
\BIBdecl

\bibitem{fetterer_sea_1994}
F.~M. Fetterer, D.~Gineris, and R.~Kwok, ``Sea ice type maps from alaska
  synthetic aperture radar facility imagery: {An} assessment,'' \emph{J.
  Geophys. Res. Oceans}, vol.~99, no. C11, pp. 22\,443--22\,458, 1994.

\bibitem{bovolo_hierarchical_2013}
F.~Bovolo, C.~Marin, and L.~Bruzzone, ``A hierarchical approach to change
  detection in very high resolution {SAR} images for surveillance
  applications,'' \emph{{IEEE} Trans. Geosci. Remote Sens.}, vol.~51, no.~4,
  pp. 2042--2054, Apr. 2013.

\bibitem{touzi_review_2002}
R.~Touzi, ``A review of speckle filtering in the context of estimation
  theory,'' \emph{{IEEE} Trans. Geosci. Remote Sens.}, vol.~40, no.~11, pp.
  2392--2404, Nov. 2002.

\bibitem{gao_statistical_2010}
G.~Gao, ``Statistical modeling of {SAR} images: {A} survey,'' \emph{Sensors},
  vol.~10, no.~1, pp. 775--795, Jan. 2010.

\bibitem{argenti_tutorial_2013}
F.~Argenti, A.~Lapini, T.~Bianchi, and L.~Alparone, ``A tutorial on speckle
  reduction in synthetic aperture radar images,'' \emph{IEEE Geosci. Remote
  Sens. Mag.}, vol.~1, no.~3, pp. 6--35, Sep. 2013.

\bibitem{lee_digital_1980}
J.-S. Lee, ``Digital image enhancement and noise filtering by use of local
  statistics,'' \emph{{IEEE} Trans. Pattern Anal. Mach. Intell.}, vol. PAMI-2,
  no.~2, pp. 165--168, Mar. 1980.

\bibitem{jong-sen_l_scattering-model-based_2006}
{Jong-Sen L}, M.~Grunes, D.~Schuler, E.~Pottier, and L.~Ferro-Famil,
  ``Scattering-model-based speckle filtering of polarimetric {SAR} data,''
  \emph{{IEEE} Trans. Geosci. Remote Sens.}, vol.~44, no.~1, pp. 176--187, Jan.
  2006.

\bibitem{frost_model_1982}
V.~S. Frost, J.~A. Stiles, K.~S. Shanmugan, and J.~C. Holtzman, ``A model for
  radar images and its application to adaptive digital filtering of
  multiplicative noise,'' \emph{{IEEE} Trans. Pattern Anal. Mach. Intell.},
  vol. PAMI-4, no.~2, pp. 157--166, Mar. 1982.

\bibitem{kuan_adaptive_1987}
D.~Kuan, A.~Sawchuk, T.~Strand, and P.~Chavel, ``Adaptive restoration of images
  with speckle,'' \emph{{IEEE} Trans. Acoust.}, vol.~35, no.~3, pp. 373--383,
  Mar. 1987.

\bibitem{lopes_structure_1993}
A.~Lopes, E.~Nezry, R.~Touzi, and H.~Laur, ``Structure detection and
  statistical adaptive speckle filtering in {SAR} images,'' \emph{Int. J.
  Remote Sens.}, vol.~14, no.~9, pp. 1735--1758, Jun. 1993.

\bibitem{chen_nonlocal_2011}
J.~Chen, Y.~Chen, W.~An, Y.~Cui, and J.~Yang, ``Nonlocal filtering for
  polarimetric {SAR} data: {A} pretest approach,'' \emph{IEEE Trans. Geosci.
  Remote Sensing}, vol.~49, no.~5, pp. 1744--1754, May 2011.

\bibitem{hua_zhong_robust_2014}
{Hua Zhong}, {Jingjing Zhang}, and {Ganchao Liu}, ``Robust polarimetric {SAR}
  despeckling based on nonlocal means and distributed lee filter,'' \emph{IEEE
  Trans. Geosci. Remote Sensing}, vol.~52, no.~7, pp. 4198--4210, Jul. 2014.

\bibitem{liu_nonlocal_2014}
G.~Liu and H.~Zhong, ``Nonlocal means filter for polarimetric {SAR} data
  despeckling based on discriminative similarity measure,'' \emph{{IEEE}
  Geosci. Remote Sens. Lett.}, vol.~11, no.~2, pp. 514--518, Feb. 2014.

\bibitem{deledalle_nl_sar_2015}
C.-A. Deledalle, L.~Denis, F.~Tupin, A.~Reigber, and M.~Jager, ``{NL}-{SAR}:
  {A} unified nonlocal framework for resolution-preserving {Pol}{In}{SAR}
  denoising,'' \emph{IEEE Trans. Geosci. Remote Sens.}, vol.~53, no.~4, pp.
  2021--2038, Apr. 2015.

\bibitem{dabov_image_2007}
K.~Dabov, A.~Foi, V.~Katkovnik, and K.~Egiazarian, ``Image denoising by sparse
  3-{D} transform-domain collaborative filtering,'' \emph{IEEE Trans. on Image
  Process.}, vol.~16, no.~8, pp. 2080--2095, Aug. 2007.

\bibitem{parrilli_nonlocal_2012}
S.~Parrilli, M.~Poderico, C.~V. Angelino, and L.~Verdoliva, ``A nonlocal {SAR}
  image denoising algorithm based on {LLMMSE} wavelet shrinkage,'' \emph{{IEEE}
  Trans. Geosci. Remote Sens.}, vol.~50, no.~2, pp. 606--616, Feb. 2012.

\bibitem{lecun_deep_2015}
Y.~LeCun, Y.~Bengio, and G.~Hinton, ``Deep learning,'' \emph{Nature}, vol. 521,
  no. 7553, pp. 436--444, May 2015.

\bibitem{wang_sar_2017}
P.~Wang, H.~Zhang, and V.~M. Patel, ``{SAR} image despeckling using a
  convolutional neural network,'' \emph{{IEEE} Signal Process. Lett.}, vol.~24,
  no.~12, pp. 1763--1767, Dec. 2017.

\bibitem{gu_residual_2017}
F.~Gu, H.~Zhang, C.~Wang, and B.~Zhang, ``Residual encoder-decoder network
  introduced for multisource {SAR} image despeckling,'' in \emph{Proc. IEEE SAR
  Big Data Era, Models, Methods Appl. (BIGSARDATA)}, Beijing, China, Nov. 2017,
  pp. 1--5.

\bibitem{chierchia_sar_2017}
G.~Chierchia, D.~Cozzolino, G.~Poggi, and L.~Verdoliva, ``{SAR} image
  despeckling through convolutional neural networks,'' in \emph{{Proc. {IEEE}
  Int. Geosci. Remote Sens. Symp. (IGARSS)}}, Fort Worth, TX, Jul. 2017, pp.
  5438--5441.

\bibitem{kim_despeckling_2018}
M.~Kim, J.~Lee, and J.~Jeong, ``A despeckling method using stationary wavelet
  transform and convolutional neural network,'' in \emph{Proc. IEEE Int. Work.
  Adv. Image Technol. (IWAIT)}, Chiang Mai, Jan. 2018, pp. 1--4.

\bibitem{zhang_learning_2018}
Q.~Zhang, Q.~Yuan, J.~Li, Z.~Yang, and X.~Ma, ``Learning a dilated residual
  network for {SAR} image despeckling,'' \emph{Remote Sens.}, vol.~10, no.~2,
  p. 196, Jan. 2018.

\bibitem{tang_sar_2018}
X.~Tang, L.~Zhang, and X.~Ding, ``{SAR} image despeckling with a multilayer
  perceptron neural network,'' \emph{Int. J. Digit. Earth}, pp. 1--21, Mar.
  2018.

\bibitem{gui_sar_2018}
Y.~Gui, L.~Xue, and X.~Li, ``{SAR} image despeckling using a dilated densely
  connected network,'' \emph{Remote Sens. Lett.}, vol.~9, no.~9, pp. 857--866,
  Sep. 2018.

\bibitem{Lehtinen2018Noise2NoiseLI}
J.~Lehtinen, J.~Munkberg, J.~Hasselgren, S.~Laine, T.~Karras, M.~Aittala, and
  T.~Aila, ``{Noise2Noise}: {Learning} image restoration without clean data,''
  in \emph{Proc. Int. Conf. Mach. Learn. (ICML)}, Stockholm, Sweden, Jul. 2018,
  pp. 2965--2974.

\bibitem{lee1989noise}
J.~Lee and K.~Hoppel, ``Noise modeling and estimation of remotely-sensed
  images,'' in \emph{Proc. IEEE Can. Symp. Remote Sens. Geosci. Remote Sens.
  Symp.}, vol.~2, 1989, pp. 1005--1008.

\bibitem{he_delving_2015}
K.~He, X.~Zhang, S.~Ren, and J.~Sun, ``Delving deep into rectifiers:
  {Surpassing} human-level performance on {ImageNet} classification,'' in
  \emph{Proc. {IEEE} Int. Conf. Comput. Vis. (ICCV) Workshop}, Santiago, Chile,
  Dec. 2015, pp. 1026--1034.

\bibitem{huang_densely_2017}
G.~Huang, Z.~Liu, L.~v.~d. Maaten, and K.~Q. Weinberger, ``Densely connected
  convolutional networks,'' in \emph{Proc. IEEE Int. Conf. Comput. Vis. Pattern
  Recognit. (CVPR)}, Honolulu, HI, Jul. 2017, pp. 2261--2269.

\bibitem{he_deep_2016}
K.~He, X.~Zhang, S.~Ren, and J.~Sun, ``Deep residual learning for image
  recognition,'' in \emph{Proc. IEEE Int. Conf. Comput. Vis. Pattern Recognit.
  (CVPR)}, Las Vegas, NV, USA, Jun. 2016, pp. 770--778.

\bibitem{ioffe_batch_nodate}
S.~Ioffe and C.~Szegedy, ``Batch normalization: {Accelerating} deep network
  training by reducing internal covariate shift,'' in \emph{Proc. Int. Conf.
  Mach. Learn. (ICML)}, Lille, France, Jul. 2015, pp. 448--456.

\bibitem{lim_enhanced_2017}
B.~Lim, S.~Son, H.~Kim, S.~Nah, and K.~M. Lee, ``Enhanced deep residual
  networks for single image super-resolution,'' in \emph{Proc. IEEE Int. Conf.
  Comput. Vis. Pattern Recognit. (CVPR) Workshop}, Honolulu, HI, USA, Jul.
  2017, pp. 1132--1140.

\bibitem{nah_deep_2017}
S.~Nah, T.~H. Kim, and K.~M. Lee, ``Deep multi-scale convolutional neural
  network for dynamic scene deblurring,'' in \emph{Proc. IEEE Int. Conf.
  Comput. Vis. Pattern Recognit. (CVPR) Workshop}, Honolulu, HI, Jul. 2017, pp.
  257--265.

\bibitem{yu_multi-scale_2015}
F.~Yu and V.~Koltun, ``Multi-scale context aggregation by dilated
  convolutions,'' \emph{arXiv:1511.07122}, Nov. 2015.

\bibitem{ILSVRC15}
O.~Russakovsky, J.~Deng, H.~Su, J.~Krause, S.~Satheesh, S.~Ma, Z.~Huang,
  A.~Karpathy, A.~Khosla, M.~Bernstein, A.~C. Berg, and L.~Fei-Fei, ``{ImageNet
  Large Scale Visual Recognition Challenge},'' \emph{Int. J. Comput. Vis.},
  vol. 115, no.~3, pp. 211--252, 2015.

\bibitem{kingma_adam_2014}
D.~P. Kingma and J.~Ba, ``Adam: {A} method for stochastic optimization,''
  \emph{arXiv:1412.6980}, Dec. 2014.

\bibitem{paszke2017automatic}
A.~Paszke, S.~Gross, S.~Chintala, G.~Chanan, E.~Yang, Z.~DeVito, Z.~Lin,
  A.~Desmaison, L.~Antiga, and A.~Lerer, ``Automatic differentiation in
  pytorch,'' in \emph{Proc. Adv. Neural Inf. Process. Syst. (NeurIPS)
  Workshop}, Long Beach, CA, USA, Oct. 2017.

\bibitem{wang_image_2004}
Z.~Wang, A.~Bovik, H.~Sheikh, and E.~Simoncelli, ``Image quality assessment:
  {From} error visibility to structural similarity,'' \emph{IEEE Trans. on
  Image Process.}, vol.~13, no.~4, pp. 600--612, Apr. 2004.

\bibitem{di_martino_benchmarking_2014}
G.~Di~Martino, M.~Poderico, G.~Poggi, D.~Riccio, and L.~Verdoliva,
  ``Benchmarking framework for {SAR} despeckling,'' \emph{{IEEE} Trans. Geosci.
  Remote Sens.}, vol.~52, no.~3, pp. 1596--1615, Mar. 2014.

\bibitem{ma_review_2018}
X.~Ma, P.~Wu, Y.~Wu, and H.~Shen, ``A review on recent developments in fully
  polarimetric {SAR} image despeckling,'' \emph{{IEEE} J. Sel. Top. Appl. Earth
  Obs. Remote Sens.}, vol.~11, no.~3, pp. 743--758, Mar. 2018.

\bibitem{oliver2004understanding}
C.~Oliver and S.~Quegan, \emph{Understanding synthetic aperture radar
  images}.\hskip 1em plus 0.5em minus 0.4em\relax Raleigh, NC, USA: SciTech,
  2004.

\bibitem{xiangli_nie_variational_2015}
X.~Nie, H.~Qiao, and B.~Zhang, ``A variational model for {PolSAR} data speckle
  reduction based on the wishart distribution,'' \emph{IEEE Trans. on Image
  Process.}, vol.~24, no.~4, pp. 1209--1222, Apr. 2015.

\bibitem{deledalle_iterative_2009}
C.-A. Deledalle, L.~Denis, and F.~Tupin, ``Iterative weighted maximum
  likelihood denoising with probabilistic patch-based weights,'' \emph{{IEEE}
  Geosci. Remote Sens. Lett.}, vol.~18, no.~12, pp. 2661--2672, Dec. 2009.

\bibitem{cozzolino_fast_2014}
D.~Cozzolino, S.~Parrilli, G.~Scarpa, G.~Poggi, and L.~Verdoliva, ``Fast
  adaptive nonlocal {SAR} despeckling,'' \emph{IEEE Geosci. Remote Sens.
  Lett.}, vol.~11, no.~2, pp. 524--528, Feb. 2014.

\bibitem{yang_bag--visual-words_2010}
Y.~Yang and S.~Newsam, ``Bag-of-visual-words and spatial extensions for
  land-use classification,'' in \emph{Proc. ACM SIGSPATIAL Int. Conf. Adv. Inf.
  (ACM GIS)}, San Jose, California, Nov. 2010, p. 270.

\end{thebibliography}
\end{document}